\documentclass{optica-article}

\journal{opticajournal} % for journals or Optica Open

\articletype{Research Article}

\usepackage{lineno}
\begin{document}

\title{Real-time, chirped-pulse heterodyne detection at room-temperature with 100GHz 3dB-bandwidth mid-infrared quantum-well photodetectors}

\author{Quyang Lin,\authormark{1} Michael Hakl,\authormark{1} Sylvie Lepillet,\authormark{1} Hua Li,\authormark{2} Jean-Fran\c cois Lampin,\authormark{1} Emilien Peytavit\authormark{1} and Stefano Barbieri\authormark{1,*}} 

\address{\authormark{1}Institute of Electronics, Microelectronics and Nanotechnology, CNRS, Univ. Lille,  Univ. Polytechnique Hauts-de-France, UMR 8520, F-59000 Lille, France\\
\authormark{2}Key Laboratory of Terahertz Solid State Technology, Shanghai Institute of Microsystem and Information Technology, Chinese Academy of Sciences, 865 Changning Road, Shanghai 200050, China}

\email{\authormark{*}stefano.barbieri@iemn.fr} %% email address is required; see note below about the corresponding author designation

% use {asbstract*} to suppress the copyright line. Copyright information will be added in production

\begin{abstract*} 
\textcolor{black}{Thanks to intrinsically short electronic relaxation on the ps time scale, III-V semiconductor unipolar devices are ideal candidates for ultrahigh-speed operation at mid-infrared frequencies}. In this work, antenna-coupled, GaAs-based multi quantum-well photodetectors operating in the 10-11$\mu$m range are demonstrated, with a responsivity of 0.3A/W and a 3dB-cutoff bandwidth of 100GHz at room-temperature.
The frequency response is measured up to 220GHz: beyond 100GHz we find a roll-off dominated by the 2.5 ps-long \textcolor{black}{recombination time} of the photo-excited electrons. The potential of the detectors is illustrated by setting up an experiment where the time dependent emission frequency of a quantum cascade laser operated in pulsed mode is measured electronically and in real-time, over a frequency range $>$60GHz. By exploiting broadband electronics, and thanks to its high signal-to-noise ratio, this technique allows the acquisition, in a single-shot, of frequency-calibrated, mid-infrared molecular spectra spanning up to 100GHz and beyond, \textcolor{black}{which is particularly attractive for fast, active remote sensing applications in fields such as environmental or combustion monitoring.} 
\end{abstract*}

%%%%%%%%%%%%%%%%%%%%%%%%%%  body  %%%%%%%%%%%%%%%%%%%%%%%%%%
\section{\label{sec:level1}Introduction}
The quest for broadband photodetectors in the mid-infrared (MIR - $\lambda$=3\text{-}12 $\mu$m), with radio-frequency (RF) bandwidths in the tens of GHz or more, has gained momentum since the end of the 80s with the advent of unipolar devices based on intersubband (ISB) transitions in III-V semiconductor heterostructures (GaAs/AlGaAs and InGaAs/InAlAs) ~\cite{Liu2007}. Thanks to ultrafast electronic non-radiative lifetimes, these structures  offer intrinsic response times in the ps range, potentially leading to RF-bandwiths of tens of GHz, provided that the detector RC time constant is short enough ~\cite{Ehret1997,Grant2006,Dougakiuchi2021,Hillbrand2021,Hakl2021,Quinchard2022}. In this respect, the recent exploitation of metallic antennas of micrometric size to in-couple the impinging mid-IR radiation to the semiconductor heterostructure active region, has opened new perspectives by allowing to shrink the detectors area, without compromising the light collection efficiency ~\cite{Palaferri2016,Palaferri2018}. On the one hand, compared to standard detectors based on so-called \textquotedblleft mesa" geometry, this allows reducing the detector's dark current without affecting the responsivity. The other advantage is a reduction of the RC time constant, which can be exploited to increase the device speed~\cite{Hakl2021}. 

In the first part of this work we have pushed forward the study and optimisation of antenna-coupled MIR quantum-well infrared photodetectors (QWIPs), in order to improve their performance both in terms of responsivity and bandwidth, and, at the same time, to try assessing experimentally what are their limiting factors. To this end we have fabricated and characterised experimentally three sets of GaAs/AlGaAs-based QWIPs, based on two-dimensional matrices of metallic patch-antennas, and measured their frequency response at room-temperature in the 0-110GHz and 140GHz-220GHz frequency bands. Depending on the number of antenna elements, we find that the latter remains within 3dB up to ~100GHz (3$\times$3 and 2$\times$2 matrices), the broadest bandwidth reported to date for photodetectors based on ISB transitions. At higher frequencies we find a roll-off between 7 and 9dB/octave. By fitting the frequency response with the help of a small-signal circuit model that we extract from impedance measurements, we conclude unequivocally that the high frequency roll-off is limited by the intrinsic carrier's capture time, of $\sim2.5$ps. 

By optimizing the QWIPs design, a maximum responsivity of ~0.3 A/W is obtained at 10.3$\mu$m wavelength, a value significantly larger than what previously reported for patch-antenna QWIPs at 300K ($\sim$ 0.15-0.2A/W) \cite{Palaferri2016,Hakl2021}. The responsivity decreases with increasing incident optical power, a fact that we attribute to optical saturation of the ISB transition\cite{Vodopyanov1997,Jeannin2023}. The corresponding saturation intensity, of only a few tens of kW/cm$^2$, is consistent with the fact that the antennas allow to obtain a radiation collection area that is larger than the physical area of the detector \cite{Jeannin2020-2}. 

Applications of ultrafast QWIPs are only at their early stage, with many exciting developments in disparate fields, such as free-space communications \cite{Flannigan2022,Pang2017,Bonazzi2022,Wang2021,Didier2022,Didier2023}, gas sensing and spectroscopy \cite{Weidmann2007,Wang2014,Macleod2015,Diaz2016,Goldenstein2016,Asselin2017,Kawai2020,Dougakiuchi2021-2}, metrology \cite{Sow2014,Argence2015}, ultrafast physics \cite{Pires2015}, and astrophysics \cite{Hale2000,Sonnabend2013,Bourdarot2020}. In the second part of this work, to assess the potential of our QWIPs for fast sensing/spectroscopy applications, we have used them to detect the heterodyne beating between a quantum cascade laser (QCL) operated in pulsed mode and another one driven in continuous wave (CW). In this way, with the help of a fast oscilloscope, we show that it is possible to measure in real-time the frequency down-chirp resulting from the thermal transient of the pulsed QCL, spanning a range of more than 60GHz. By allowing the acquisition of frequency-calibrated gas spectra with a high signal-to-noise ratio in a single-shot, over timescales from tens of ns to ms, this technique appears to be particularly promising for active remote sensing and laser ranging applications.

\section{\label{sec:level2} Results}
\subsection{\label{Resp} Spectral characterisation and device responsivity}

\begin{figure}
			\includegraphics[width=\linewidth]{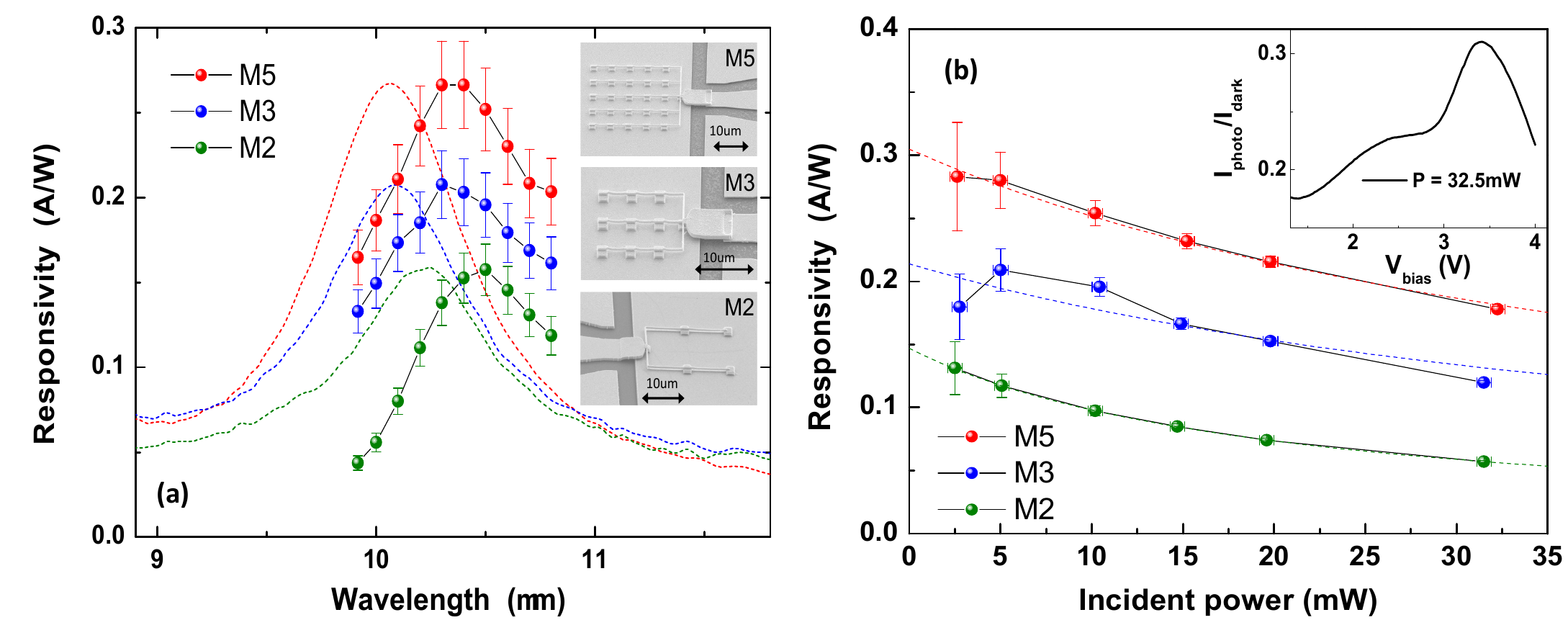}
				\caption{(a)  Room-temperature responsivity  \emph{vs} wavelength of the QWIPs studied in this work, measured with an extended-cavity QCL (dots). The incident power is of 4.3mW. For comparison, the absorption spectrum of each device measured with an FTIR spectrometer, normalised to its peak responsivity is also reported (dashed lines). The devices are labelled respectively M5, M3 and M2, and their SEM pictures are shown on the right. For all of them the patches consist of metallic squares of lateral side $s = 1.8\mu$m. In the M5 and M3 devices, individual patches are separated by a period $p = 5\mu$m, while for device M2 we used $p = 10\mu$m. As shown in the pictures, patches are electrically connected together and to a $50\Omega$ integrated coplanar line by a suspended gold wires of $\sim 150$nm diameter (only the first part of the coplanar line is visible in the SEM images). (b) Responsivity \emph{vs} incident power at $\lambda = 10.3\mu$m for the 3 devices studied, measured at 300K. The applied biases are 3.9V (devices M5 and M3) and 3.75V (device M2). The error bars take into account the uncertainty in the measurement of the incident power ($\sim \pm 0.4mW$). \textcolor{black}{The dashed lines are the fits of the responsivities using the function $R = R_{0}/(1+P_{inc}/P_{sat})$ (see the main text).} Inset. Device M5: ratio between the photocurrent at 32.3mW incident power and the dark current.} 
				\label{fig:Responsivity}
\end{figure}

The QWIP semiconductor active region consists of six, 6nm-thick, $n$-doped GaAs quantum wells (QWs) separated by 40nm-thick, undoped Al$_{0.2}$Ga$_{0.8}$As barriers, yielding a nominal bound to quasi-bound ISB transition energy of $\sim 115$meV ($\lambda \sim 10.8\mu$m). Details on the heterostructure layers and device fabrication are given in \textcolor{blue}{Methods}. The final device geometry is a matrix of square metallic (Ti/Au) patches of side $s$ and separated by a period $p$. Around each patch the semiconductor is etched down to a bottom metallic ground-plane. As shown in the SEM pictures in Fig.~\ref{fig:Responsivity}(a) the patches are electrically connected together, and to a $50\Omega$ microwave coplanar line for RF extraction, by $\sim 150$nm wide, Ti/Au wire air-bridges. 

In this work we have studied matrices with different number of patches in order to probe the effect on the photodetectors RC time constant. The devices are based on a 5$\times$5 and a 3$\times$3 matrix of period $p = 5\mu$m, and a 2$\times$2 matrix of period $p = 10\mu$m, that we label M5, M3 and M2 respectively.  For all the devices $s = 1.8\mu$m. This parameter defines the frequency of the fundamental TM$_{010}$ mode of a single resonator, the one we are interested in, which is, essentially, a $\lambda/2$ Fabry-Perot mode oscillating in the plane of the patches, perpendicularly to the connecting wire bridges \cite{Balanis2005,Todorov2010}.  The TM$_{100}$ mode oscillating in the orthogonal direction is instead perturbed by the wire bridges (despite their small size), leading to a lower overlap with the QWIP active region, and therefore a weaker absorption \cite{Hakl2021}. 

For a given $s$, changing the periodicity $p$ affects the radiation collection area of each individual patch in the array \cite{Jeannin2020,Rodriguez2020}. The experimental characterisation and optimisation of the optical absorption of the patch-antenna arrays, made with the help of a MIR microscope coupled to a Fourier transform (FTIR) spectrometer, was carried out over a large number of matrices by varying $s$ and $p$. The main results are summarised in \textcolor{blue}{Supplement 1}. In the case where the optical excitation area is smaller than the surface of the matrix (i.e. \textquotedblleft infinite" matrix approximation), for the TM$_{010}$ mode we find peak absorptions at $\sim10.5\mu$m (i.e. virtually coincident with the nominal wavelength of the ISB transition) of 96\% and 40\% for $p = 5\mu$m and $p = 10\mu$m respectively. In the former case we are therefore very close to so-called \textquotedblleft critical" coupling (100\% peak optical absorption). The reason why we choose $p=10\mu$m for device M2, is the results of a compromise between the need to keep a sizeable antenna collection area while having a reasonable spatial overlap with the waist of the focused QCLs used throughout this work, of approximately $25 \mu$m diameter (see below).

The room-temperature responsivity of the devices $vs$ wavelength in the range 9.9$\mu$m-10.8$\mu$m, obtained with an extended cavity (EC) QCL polarized perpendicularly to the connecting wires is reported in Fig.~\ref{fig:Responsivity}(a) (dots), for an incident power of 4.3mW. The QCL beam was focused with an AR coated aspheric chalcogenide-glass lens (NA = 0.56; 5 mm focal length), yielding a waist diameter of $\sim25 \mu$m, that we measured with a razor blade. We obtain a maximum responsivity close to 0.3A/W at $10.3\mu$m for device M5. As expected the responsivity is reduced by decreasing the number of patches. Indeed the waist area roughly matches that of a 5$\times$5 matrix. As a consequence, especially for devices M3 and M2, part of the incident radiation is directly reflected by the metallic ground-plane. 

The dashed lines in Fig.~\ref{fig:Responsivity}(a) represent the experimental optical absorption for each device, normalised to its peak responsivity (\textcolor{blue}{Supplement 1}). The observed systematic red shift between the  peak absorption and peak responsivity is a consequence of the fact that the QWIP ISB transition energy is not perfectly coincident with the energy of the TM$_{010}$ cavity mode. The QWIP absorption can be computed analytically using Coupled Mode Theory (CMT) \cite{Haus2004,Jeannin2020,Jeannin2020-2}: for device M5 we find a good agreement with the experimental absorption spectrum assuming an ISB transition energy $E_{isb}=115$ meV and a cavity mode energy of $E_{cav}=122.5$meV (\textcolor{blue}{Supplement 1)}. This gives an external quantum efficiency of $\sim 15\%$ for detector M5. We note that in the case where the ISB transition energy was perfectly coincident with that of the cavity mode ($E_{isb}=E_{cav} = 122.4$meV), this value would raise to $\sim 25\%$, with a corresponding peak responsivity of $\sim 0.5$A/W. 

As reported in Fig.~\ref{fig:Responsivity}(b), the responsivity of the devices measured at $\lambda = 10.3\mu$m displays a sizeable decrease (up to $\sim 40-60\%$ depending on the number of patches) with increasing power. In Ref.\cite{Jeannin2020-2} it was shown that the optical saturation intensity of an ISB transition system can be strongly reduced if the latter is embedded inside an optical cavity of sub-wavelength volume, as is the case here. Using CMT, we compute a saturation intensity for our patch-antenna $I_{sat} \sim 35$kW/cm$^2$ at $\lambda = 10.3\mu$m. To estimate the corresponding incident saturation power, $P_{sat}$, we must take into account the fact that each patch-antenna in the array collects photons on a surface larger than its physical area. As a result, at critical coupling, the incident saturation intensity is obtained by multiplying $I_{sat}$ by the factor $s^{2}/p^{2}$ (\textcolor{blue}{Supplement 1}). Considering a waist diameter of $\sim 25\mu$m, and taking into account the different peak absorptions of each detector we finally obtain $P_{sat} \sim$ 30mW, 45mW, and 20mW for QWIPs M5, M3 and M2 respectively. \textcolor{black}{The dashed lines in Fig.~\ref{fig:Responsivity}(b) represent the fits of the responsivities using the function $R = R_{0}/(1+P_{inc}/P_{sat})$, where $P_{inc}$ is the incident power and $R_{0}$ and $P_{sat}$ are used as fitting parameters ($R_{0}$ is the responsivity at low incident power) \cite{Jeannin2020-2}. From the fits we obtain $P_{sat} = 47 \pm 3$mW, $50 \pm 20$mW and  $20 \pm 0.1$mW for QWIPs M5, M3 and M2 respectively, in fairly good agreement with the computed values.}

\subsection{\label{sec:level3} Frequency response}

The experimental setup for the measurement of the QWIPs frequency response is based on the heterodyne mixing of a DFB QCL emitting at $\sim 10.3\mu$m with an EC QCL (the same used for Fig.~\ref{fig:Responsivity}(a)). Both lasers are operated in CW, and a MIR isolator is used to minimise optical feedback. As a consequence the incident radiation is linearly polarised along the diagonal of the square patches, resulting into a $\sim 50\%$ drop of absorption compared to Fig.~\ref{fig:Responsivity}. The  incident powers on the QWIPs are $P_{1} = 13$mW and $P_{2}=17.5$mW from the EC and DFB QCLs respectively. To avoid parasitic effects due to wire-bonding/packaging, the measurement of the heterodyne signal, oscillating at the difference between the emission frequencies of the two QCLs, is done directly on-wafer by positioning two sets of coplanar probes at the the edge of the integrated $50\Omega$ coplanar line, followed by a bias-tee and a calibrated power meter covering respectively the 0-110GHz and 140GHz-220GHz frequency bands.  

\begin{figure}
			\includegraphics[width=\linewidth]{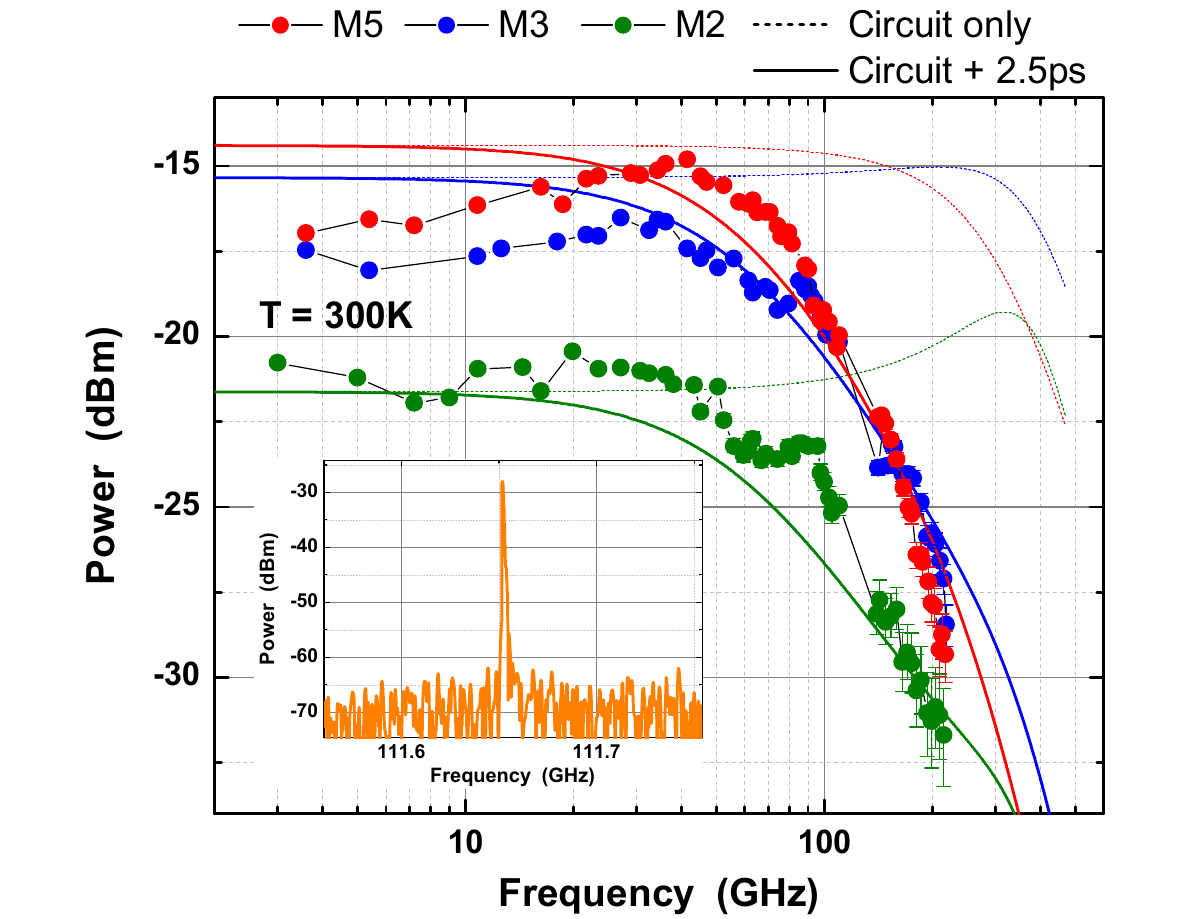}
				\caption{(a)  Room-temperature experimental frequency response of devices M5 (red dots), M3 (blue dots) and M2 (green dots) in the frequency bands 0-110GHz and 140-220GHz. The measurement are obtained at $\lambda \simeq 10.3\mu$m by heterodyne mixing two single-mode QCLs, and the experimental values are corrected by the attenuation of the bias-tees and coplanar probes, measured with a VNA. All data was recorded without the use of any amplification stage. The incident MIR radiation is linearly polarised at 45deg with respect to the metallic wires connecting the patches, and the incident powers are 13mW and 17.5mW. The corresponding $dc$ photocurent are $4.1$mA, $2.8$mA, and $1.25$mA for devices M5, M3 and M2 respectively.  The dashed lines are the computed  electrical frequency responses obtained from the small-signal equivalent circuit (\textcolor{blue}{Supplement 1}, Fig. S4). The solid lines include the intrinsic ISB frequency response with a carrier capture time of 2.5ps. Inset. Example of heterodyne beatnote close to 110GHz, recorded with a spectrum analyser using the M3 photodetector. The spectrum is not corrected by the attenuation through the 110GHz probe and bias-tee.}
				\label{fig:Response}
\end{figure}

In Fig.~\ref{fig:Response} we report representative experimental frequency response functions for devices M5, M3 and M2, obtained by sweeping the emission frequency of the EC QCL using the external grating, while the DFB QCL is kept at constant current. The devices are biased at 3.8V (M5), 3.85V (M3) and 4V (M2), corresponding to the maximum generated photocurrents (\textcolor{blue}{Supplement 1}). The experimental power values are corrected by the attenuation of the bias-tees and coplanar probes, measured with a Vector Network Analyser (VNA). We obtain 3dB cutoffs of $\sim 90$GHz for device M5 and of $\sim 100$GHz for devices M3 and M2 (the cutoffs are defined relatively to the peak response). These are the largest bandwidths reported to date in the literature for unipolar MIR photodetectors and, more generally, for MIR photodetectors. Beyond the 3dB cutoff the response drops by approximately 8dB/octave. 

The frequency response of the photodetector is essentially the product of two transfer functions, the first including the electrical response, while the second one takes into account the intrinsic response time of the photo-excited electrons \cite{Liu2007}. To obtain the electrical response functions of the devices studied, we first measured their impedance and then used the latter to derive an equivalent small-signal circuit model (\textcolor{blue}{Supplement 1}). The frequency response can then be obtained by computing the average power, $P_{L}(\omega)$ dissipated in the  $50\Omega$ input impedance of the power meter, where $\omega$ is the difference frequency between the two QCLs, and considering an $ac$ current source term of amplitude $I_{s}$ proportional to the total $dc$ photocurrent generated by the two QCLs (\textcolor{blue}{Methods}). The dashed lines in Fig.~\ref{fig:Response} are the so-obtained electrical frequency responses. Clearly, the predicted cutoff frequencies are much larger than those observed experimentally, i.e. the response time of our photodetectors is not limited by the electrical time constant but rather by the intrinsic response time of the ISB system, which can be taken into account by multiplying the electrical transfer function by the term $[1+(\omega \tau)^{2}]^{-1/2}$, where $\tau$ represents the shortest between the carriers capture time and transit time \cite{Liu2007}. 
The best agreement with the experimental frequency responses is shown by the solid lines in Fig.~\ref{fig:Response}, obtained with $\tau = 2.5$ps, that we identify with the carriers capture time. Indeed, under the experimentally applied biases we estimate a drift velocity at room temperature of $2-3\times 10^{6}$ cm/s, yielding a transit time of $\sim 10$ps  \cite{Hakl2021,Palaferri2018}.

\subsection{\label{sec:level4} Heterodyne frequency-chirp spectroscopy} 

It is well-known that driving a QCL in pulsed mode generates a down-chirp of the emission frequency of thermal origin, that can reach up to several tens of GHz. This effect can be exploited to detect in real time different gas species for applications in environmental and combustion monitoring, plasma diagnostic, or high-resolution spectroscopy \cite{Namjou1998,McCulloch2003,Normand2003,Grouiez2010,Tombez2013,Michael2005,Welzel2010}. 

\begin{figure}
\includegraphics[width=\linewidth]{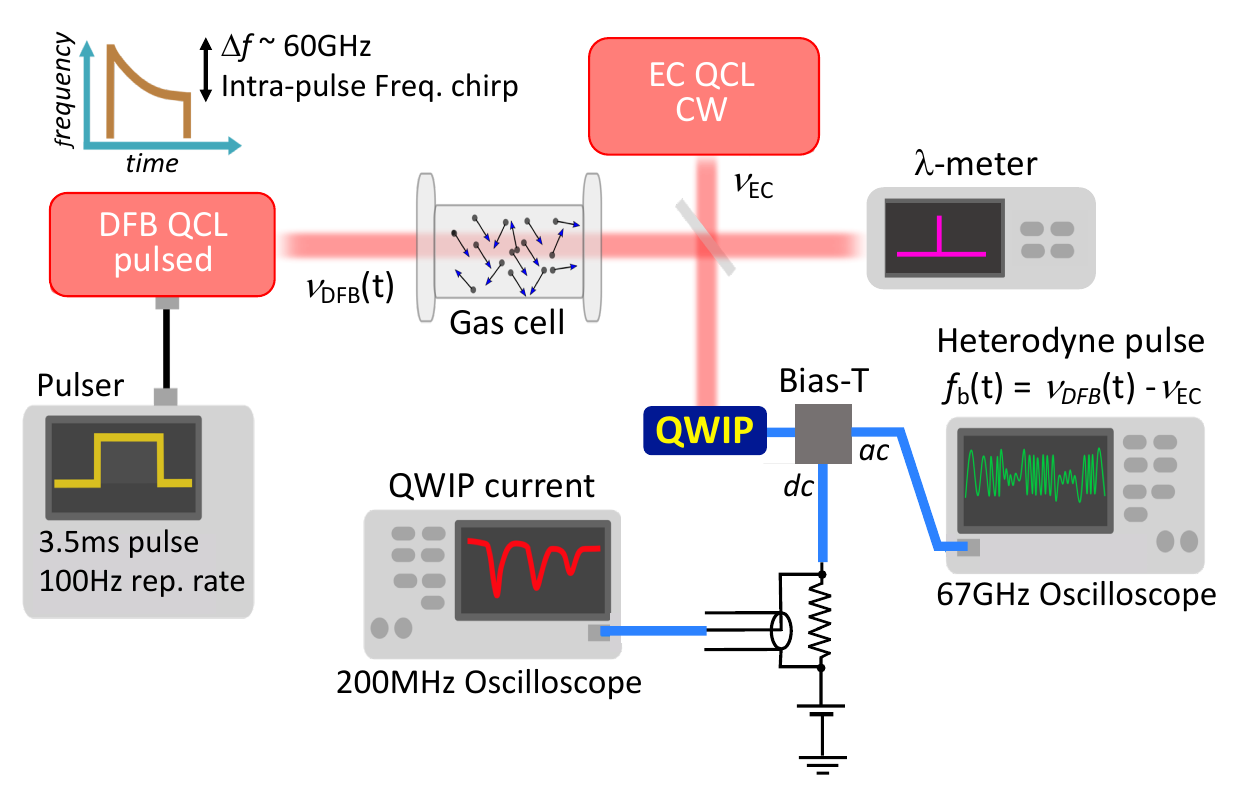}
\caption{Schematic of the HFCS experiment (see main text for more details). A DFB QCL, with emission frequency $\nu_{DFB}$(t), is driven in pulsed mode, producing a frequency down-chirp of approximately 60GHz.  Its beam is transmitted through a gas cell containing NH$_{3}$ and focused on the QWIP. A tunable EC QCL is driven in CW and provides the local oscillator for heterodyne detection. Its absolute frequency, $\nu_{EC}$, is monitored with a $\lambda$-meter. 
The QWIP, in series with a 34$\Omega$ resistor, is connected to a 67GHz bias-tee and biased with a $dc$ power supply. A 200MHz bandwidth oscilloscope is used to measure the voltage across the 34$\Omega$ resistor, proportional to the QWIP current. The $ac$ port of the bias-tee is connected to a 70GHz bandwidth oscilloscope, which measures in real-time the heterodyne frequency pulse, oscillating at $f_{b}(t) = \nu_{DFB}(t)-\nu_{EC}$, resulting from the mixing between the DFB and the EC QCLs.}
\label{fig:Setup}
\end{figure}
In a typical experiment, the beam emitted by a pulsed QCL is transmitted through a gas cell, then focused on a detector of sufficiently high speed to resolve the optical pulse, which is finally connected to an oscilloscope. The resulting electrical pulse will display a number of dips generated each time the QCL frequency goes across a molecular absorption line. One weak point of this technique is that the value of the QCL emission frequency at each instant of time during the pulse is not known, a fact that can be problematic, for instance for the determination of unknown transition lines. For sufficiently short driving pulses the frequency chirp is approximately linear, allowing an absolute frequency pre-calibration using a Fourier transform (FT) spectrometer \cite{Normand2003}. The generation of wider frequency spans requires instead longer driving pulses, typically ranging from tens of $\mu$s to several ms, during which the time dependence of the QCL frequency is highly non-linear, requiring the use of an etalon for real-time relative frequency calibration \cite{Grouiez2010}. 
An alternative solution to this problem is offered by the 100GHz bandwidth of our QWIP, giving the possibility to measure in real-time the relative emission frequency of a pulsed QCL through heterodyne detection. 

The schematic of the heterodyne frequency-chirp spectroscopy
(HFCS) experimental setup exploiting the same QCLs used to characterise the QWIPs frequency response is shown in Fig.~\ref{fig:Setup}. 
\begin{figure}
\includegraphics[width=\linewidth]{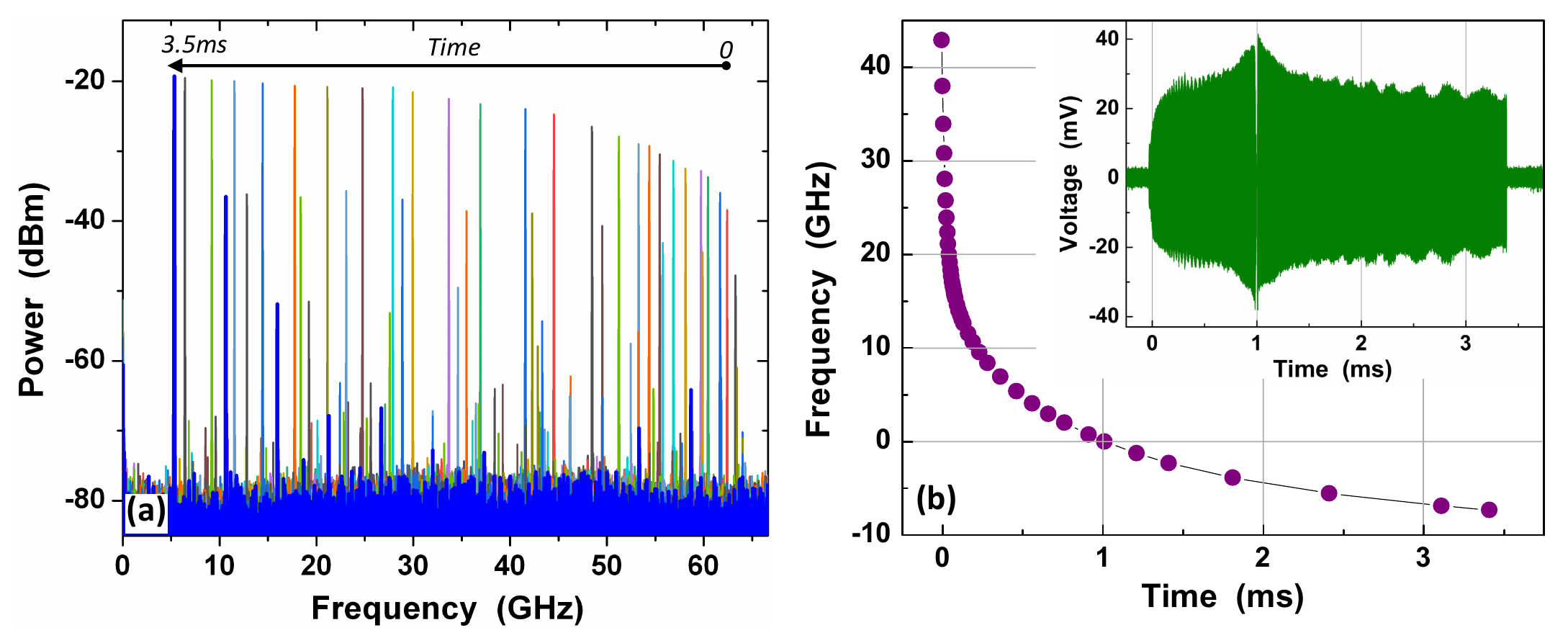}
\caption{(a) Example of heterodyne spectra obtained by computing, in real time, the FFT of a chirped pulse over a 10ns-long sliding temporal window. Time increases from right to left as schematically shown by the top arrow. For clarity, the lowest frequency spectrum is displayed in blue, showing the fundamental oscillation, close to 5GHz, and a few lower power harmonics, stemming from the QWIP I/V non-linearity and/or a circuit non-linearity (\textcolor{blue}{Supplement 1}) \cite{Peytavit2021} The chirped pulse is different from the one shown in the inset of panel (b), with the QCLs operating conditions set to obtain always a positive $f_{b}(t)$.
The pulse was recorded without gas cell, yielding $\sim 15$mW of peak and CW power incident on the QWIP. (b) Beat-frequency $vs$ time obtained from the heterodyne chirped pulse shown in the inset. The frequency was obtained by recording the same type of heterodyne spectra shown in panel (a). Inset. Single-shot chirped pulse obtained by driving the DFB QCL at 23.8$^{\circ}$C with 3.5ms-long 995mA pulses, and 100Hz repetition rate. The beam of the DFB QCL is transmitted through the empty gas cell, producing a $\sim 30\%$ power attenuation. The EC QCL is driven in CW at 1A and 17.8$^{\circ}$C yielding an emission frequency 29004.6 GHz (10.343$\mu$m). The peak and CW power incidents on the QWIP are of approximately 10mW.}
				\label{fig:ChirpedPulse}
\end{figure}
The $\sim10.3\mu$m-wavelength DFB QCL is driven in pulsed mode, with 3.5ms-long pulses and 100Hz repetition rate, producing a frequency down-chirp of approximately 60GHz (see below).  The emitted optical beam is transmitted through a \textcolor{black}{8cm-long gas cell} containing NH$_{3}$ and finally focused on a QWIP nominally identical to the M5 device of Fig.~\ref{fig:Responsivity}(a). The tunable EC QCL is driven in CW and directly focused on the QWIP to provide the local oscillator for heterodyne detection. Its absolute frequency is monitored with a Fourier transform-based $\lambda$-meter with a frequency resolution of 1GHz. As for the characterisation of the frequency response, an optical isolator (not shown in the figure) is placed before the QWIP. The QWIP is in series with a 34$\Omega$ resistor, and is connected to a 67GHz bias-tee. The $dc$ port of the latter is used to bias the QWIP with a $dc$ power supply ($\sim 4.5$V applied bias - not shown in the Figure). Simultaneously, we measure the voltage across the 34$\Omega$ resistor, proportional to the QWIP current, with the help of a 200MHz bandwidth oscilloscope. The $ac$ port of the bias-tee is connected to a 70GHz bandwidth oscilloscope, allowing to measure in real-time the heterodyne frequency pulse resulting from the mixing between the DFB and the EC QCLs. As for the heterodyne measurement of the frequency response, we note the absence of any RF amplification stage in this experimental setup. 

An example of heterodyne pulse, recorded in single-shot with the gas cell empty, is shown in the inset of Fig.~\ref{fig:ChirpedPulse}(b) (see the Figure caption for the QCLs driving conditions and the power incident on the QWIP). 
\begin{figure}
\includegraphics[width=\linewidth]{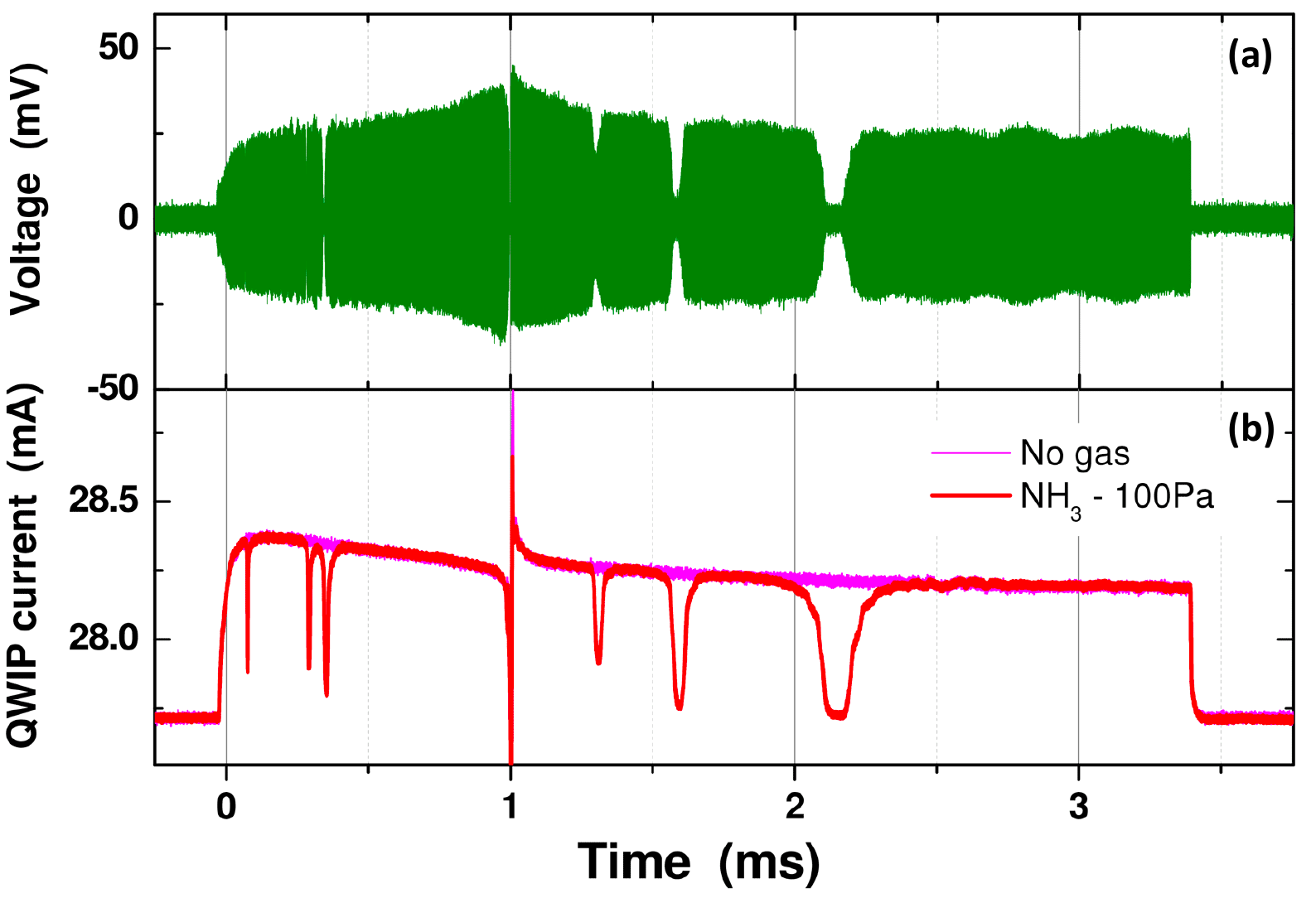}
				\caption{(a) Single-shot chirped-frequency pulse obtained by driving the QCLs under the same conditions used for Fig.~\ref{fig:ChirpedPulse}(b). The beam from the pulsed DFB QCL is transmitted through the gas cell filled with NH$_3$ at a nominal pressure of 100Pa. (b) Current pulse measured on the $dc$ port of the bias-tee (see Fig.~\ref{fig:Setup}), with the gas cell empty (purple) and filled with NH$_3$ (red).}
				\label{fig:NH3-Spectroscopy-1}
\end{figure}
\textcolor{black}{The heterodyne amplitude oscillations cannot be resolved directly using the full chirped pulse since the latter does not contain a sufficiently large number of points. The instantaneous frequency is therefore obtained by measuring, at different instants of time, single-shot, 10ns-long time traces, and by computing their Fourier transform in real time with the help of the 70GHz oscilloscope}.
This gives rise to the type of RF spectra  shown in Fig.~\ref{fig:ChirpedPulse}(a) obtained, without gas cell, from a chirped pulse different from the one shown in the inset (see caption of Fig.~\ref{fig:ChirpedPulse}). As shown by the one highlighted in blue in the Figure, each RF spectrum consists of a main peak followed by a few low power harmonics, with the former corresponding to the instantaneous beat frequency between the DFB and EC QCLs emission frequencies: $f_{b}(t) = \nu_{DFB}(t)-\nu_{EC}$. As shown by the top arrow, from 0ms to 3.5ms $f_{b}(t)$ spans approximately 60GHz. We note the high dynamic range obtained (up to 60dB) despite the fact that the chirped pulse is acquired without amplification and in single-shot. Indeed, we found that the introduction of an averaging produced a reduction of the pulse amplitude that we attribute to the frequency fluctuations of the EC QCL operating in free-running, automatically transferred to $f_{b}(t)$. This problem could be solved by locking the EC QCL to a more stable reference \cite{Sow2014,Argence2015}.

The temporal evolution of $f_{b}(t)$ is highly non-linear. This is shown in Fig.~\ref{fig:ChirpedPulse}(b), reporting the beat frequency as obtained from the chirped-pulse in the inset. The observed down-chirp is of pure thermal origin and reflects the heating of the active region due to the applied current pulse. As discussed in Ref.\cite{Tombez2013} this process involves several time constants, corresponding to joule heating diffusing through the laser active region, waveguide, substrate etc. We note that close to 1ms, $f_{b}(t)$ goes through zero, which corresponds to the point where the DFB and EC QCLs frequencies are equal.
This produces a smooth peak in the envelope of the heterodyne pulse, since as $f_{b}$ moves away from $dc$, we have an increase of the microwave propagation losses of the 1m-long, 67GHz coaxial cable connecting the $ac$ port of the bias-tee to the 70GHz oscilloscope.
Adding the emission frequency of the EC QCL measured with the $\lambda$-meter to the heterodyne frequency of Fig.~\ref{fig:ChirpedPulse}(b) provides the temporal evolution of the DFB QCL absolute emission frequency. This can then be used as a calibration for HFCS. 

The result of a proof-of-principle HFCS experiment is shown in Fig.~\ref{fig:NH3-Spectroscopy-1}, obtained by filling the gas cell with pure NH$_{3}$ at a nominal pressure of 100Pa. The top panel shows the chirped-frequency pulse, while the current pulse measured on the $dc$ port of the bias-tee is reported in the bottom panel, together with the pulse without gas for comparison. In both time-traces, several absorption dips are visible, corresponding to NH$_{3}$ absorption lines, while the spike at $\sim 1$ms in the QWIP current is an experimental artefact produced by $f_{b}(t)$ passing through 0. It is worth noting that, contrary to the chirped pulse, recorded in single-shot, the current pulse is obtained by averaging over 100 time-traces \textcolor{black}{(see \textcolor{blue}{Methods} for a comparison between the chirped pulse and the current pulse in single-shot, and for the pressure detection limit)}. 

\begin{figure}
\includegraphics[width=\linewidth]{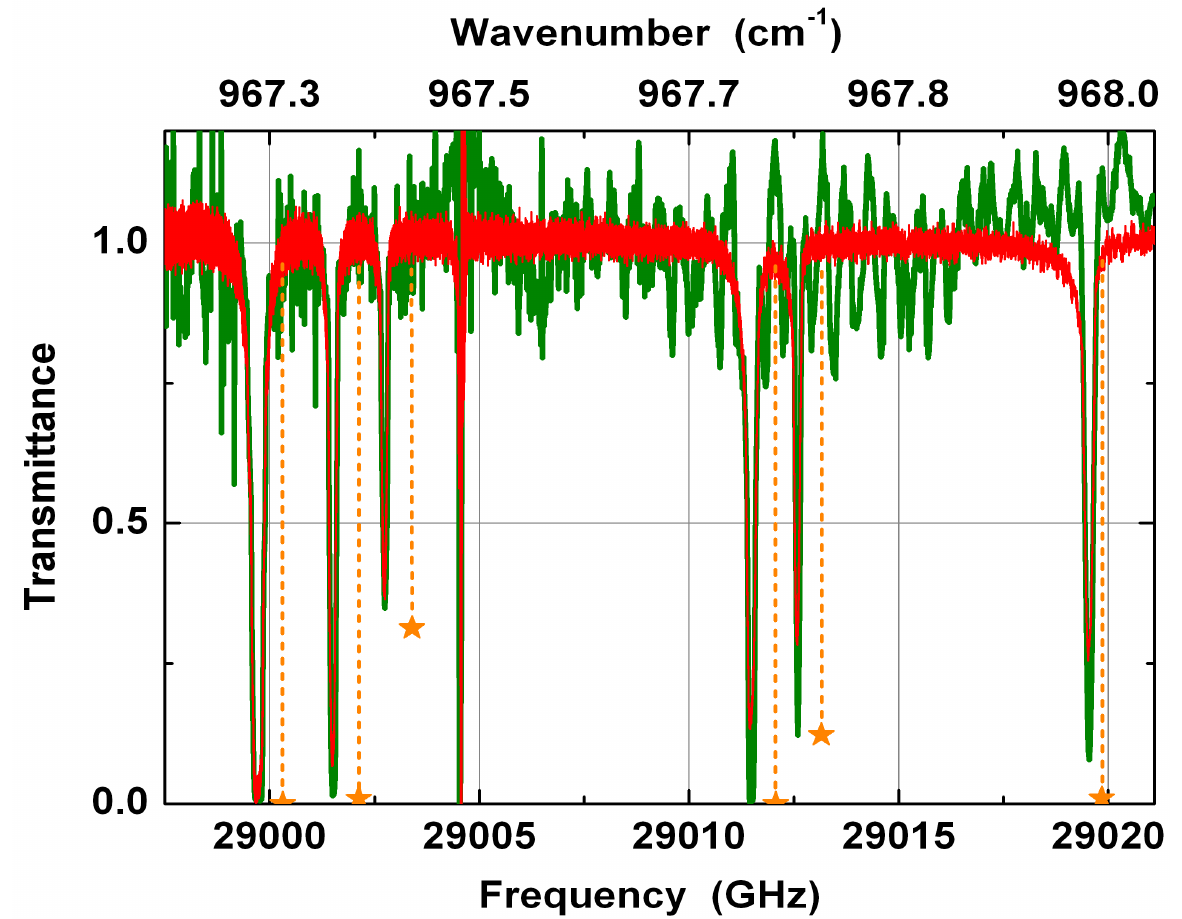}
			\caption{Green line. NH$_3$ transmission spectrum obtained from the ratio between the squares of the heterodyne pulses with and without gas (the pulse with gas is the one displayed in Fig.~\ref{fig:NH3-Spectroscopy-1}(a)). Red line. NH$_{3}$ spectrum derived from the ratio between the current pulses with and without gas of Fig.~\ref{fig:NH3-Spectroscopy-1}(b). The orange stars represent the frequency and the transmission intensities of the closest NH$_{3}$ ro-vibrational transitions, based on the HITRAN database and computed with the commercial software Spectracalc\textsuperscript{\textregistered}.}
				\label{fig:NH3-Spectroscopy-2}
\end{figure}

The solid green line in Fig.~\ref{fig:NH3-Spectroscopy-2} represents the NH$_{3}$ transmission spectrum extracted from the heterodyne pulses, where the time axis has been replaced by the absolute frequency of the chirped QCL based on the linear interpolation of the frequency $vs$ time curve displayed in Fig.~\ref{fig:ChirpedPulse}(b).  The spectrum is the result of the ratio between the squares of the voltage heterodyne pulses (proportional to the transmitted power) with and without gas (the pulse with gas is the one displayed in Fig.~\ref{fig:NH3-Spectroscopy-1}(a)). To remove the heterodyne oscillations both time traces where numerically averaged. For comparison, the red line shows the NH$_{3}$ spectrum derived from the ratio between the current pulses with and without gas of Fig.~\ref{fig:NH3-Spectroscopy-1}(b). As expected, the frequencies of the absorption lines in the two spectra are perfectly coincident. The orange stars represent the frequencies and the transmission intensities of the closest NH$_{3}$ ro-vibrational transitions, based on the HITRAN database and computed with the commercial software Spectracalc\textsuperscript{\textregistered}, \textcolor{black}{using a gas pressure of 90Pa and a 8-cm gas cell length, i.e. equal to the nominal one. The agreement with the computed line intensities is very good, considering that the difference with the nominal gas pressure of 100Pa is within the measurement error.} 
In Table I we report the HITRAN and measured frequencies, showing that for all the lines except the highest frequency one, we find a nearly constant shift of $\sim 600$MHz that is within the resolution (1GHz) of the $\lambda$-meter  used to measure the frequency of the CW QCL. The reason why the saQ(1,1) transition is shifted by only 300MHz could be due to a drift of the EC QCL during the acquisition of the chirped frequency values displayed in Fig.~\ref{fig:ChirpedPulse}(b), which were necessarily measured at different times. Further measurements would be needed to clarify this point, which is however outside the scope of this work. On this issue, it is anyway important to note that the frequency calibration procedure based on the linear interpolation of the data-points of Fig.~\ref{fig:ChirpedPulse}(b), which has been used here for illustrative reasons, is not strictly necessary. Indeed, a faster and possibly more precise way of determining the absolute frequency of a given transition line, is to directly measure the value of the chirped frequency by using a 10-ns time-window positioned right on top of corresponding transmission dip (after removing the gas if the transmission is too low).

The higher noise visible in the green spectrum compared to the red one, is partly due to slow amplitude oscillations in the heterodyne pulse, due to standing-wave effects (see Fig.4(b)) that could not be completely removed by the normalisation process. Another source of noise is due to the QCLs loosing their coherence, giving rise to short random frequency fluctuations. This problem should be removed by stabilising the two lasers sources. Finally, in Fig.~\ref{fig:NH3-Spectroscopy-2}, we observe that corresponding to the three highest frequency NH$_{3}$ transition lines,
the red spectrum shows a systematically higher transmission compared to the green one, \textcolor{black}{as well as a slight line asymmetry}. We attribute this facts to the finite transient response time of the voltage source used to bias the QWIP, effectively limiting the current rise time when the frequency of the pulsed QCL sweeps across the absorption lines (see \textcolor{blue}{Methods}).
This experimental artifact is not present on the $ac$ port of the bias-tee, where slow bias variations are filtered out, \textcolor{black}{yielding a transmission spectrum with perfectly symmetrical lines (see \textcolor{blue}{Supplement 1} for a comparison between the transmittance of all the measured transitions as obtained from the chirped pulse, with those computed with Spectracalc\textsuperscript{\textregistered})}.

\begin{table}[b]
\caption{\label{tab:table4}%
NH$_{3}$ ro-vibrational transitions. The table reports the line center frequencies, rounded to 0.1 GHz, obtained from the HITRAN database and from Fig.\ref{fig:NH3-Spectroscopy-2}(b). $\Delta f$ is their difference.}

\begin{tabular}{cccc}
Transition&&Frequency (GHz)\\
\hline
    &HITRAN&This work&$\Delta f$\\
saQ(3,3)&29000.3& 28999.7  & 0.6\\
saQ(3,2)&29002.1& 29001.5  & 0.6\\
saQ(3,1)&29003.4& 29002.7  & 0.7\\
saQ(2,2)&29012.1& 29011.5  & 0.6\\
saQ(2,1)&29013.2& 29012.6  & 0.6\\
saQ(1,1)&29019.8& 29019.5  & 0.3\\
\end{tabular}

\end{table}

\section{\label{sec:level6}Conclusions}

In this work we have demonstrated that antenna-coupled MIR unipolar quantum-well photodetectors based on ISB transitions can reach a 3dB RF bandwidth of 100GHz at room temperature, with a responsivity of $\sim 0.3$A/W at $10.3 \mu$m wavelength. By fabricating and characterising photodetectors containing different numbers of patch-antennas we have demonstrated that the high frequency cutoff is not limited by the device parasitics, but rather by the intrinsic properties of the semiconductor heterostructure itself, namely the carriers capture time, of the order of 2.5ps. 

Thanks to their ultra-broad bandwidth we believe that the demonstrated detectors are particularly appealing as heterodyne receivers for applications as diverse as MIR astronomy, light detection and ranging (LIDAR), spectroscopy or free-space communications \cite{Hale2000,Sonnabend2013,Bourdarot2020,Weidmann2007,Wang2014,Macleod2015,Diaz2016,Kawai2020,Dougakiuchi2021,Flannigan2022,Pang2017,Bonazzi2022,Didier2022}. Indeed operating these devices as direct detectors at room temperature is less attractive due to their high dark current. Instead, besides the obvious benefits of coherent detection, adopting a heterodyne configuration gives in principle the possibility to reach a detection limited by the photon noise if the local-oscillator photocurrent is larger than the thermally activated dark current. As shown in the inset of Fig.~\ref{fig:Responsivity}(b), at the actual operating wavelength of $\sim 10 \mu$m, this seems out of reach at T=300K, due to the elevated dark current and to the observed decrease of the responsivity with increasing power that we interpret as the result of a partial optical saturation. This phenomenon was never observed before in a QWIP  \cite{Vodopyanov1997,Gomez2015,Jeannin2023} and is, in a way, the drawback of coupling the ISB structure to an antenna, which permits to achieve a higher detectivity at the price of a lower saturation power \cite{Jeannin2020-2,Jeannin2023}. Although $I_{sat}$ can be increased by increasing the doping in the QWs (\textcolor{blue}{Supplement 1}), however, according to our estimates, this gain would be quickly compensated by the growth of the dark current which depends exponentially on $n_{s}$.
On the other hand,
preliminary data as a function of temperature indicate that it should be possible, with the present detector, to achieve a shot-noise limited detection in proximity of T=250K (or possibly higher in the case where the frequencies of the ISB transition and of the patch resonators were perfectly matched, see Section \ref{Resp}), which can be reached with a thermoelectric cooler. In terms of RF bandwidth, although the present 100GHz is probably enough for most applications, a possibility to improve it would be to reduce the capture time, for instance by reducing the barriers width which, at the moment is comparable to the estimated carrier's mean free path \cite{Liu2007}.  In this respect we note that an experimental study on the dependence of MIR patch-antenna QWIPs performance (e.g. responsivity, bandwidth etc) on parameters such as the active region thickness or the number of QWs is presently lacking \cite{Rodriguez2020}.

To demonstrate the potential of our detectors as heterodyne receivers we have setup a proof-of-principle experiment where the chirped-frequency emitted by a QCL driven in pulsed mode is down-converted in the microwave range through the mixing with a second QCL operated in CW. In this way it is possible to record in real-time molecular spectra spanning up to 100GHz (and beyond), limited by the bandwidth of our detector. Contrary to conventional chirped pulsed spectroscopy, our HFCS technique simplifies the absolute calibration of the chirped frequency. Most importantly it permits to achieve high SNRs ($\sim60$dB in 100MHz bandwidth with $\sim 15$mW of peak and CW power respectively from the pulsed and CW QCLs - see Fig.~\ref{fig:ChirpedPulse}(a)), which in our opinion, makes patch-antenna QWIPs particularly attractive for remote sensing applications and also free-space communications. \textcolor{black}{In particular the reported high SNR shows that the pulsed QCL beam should still be detectable after propagating through the atmosphere by several tens of km in adverse weather conditions \cite{Corrigan2009}}.

To this end we note that much higher SNRs could be reached by locking the CW QCL to a more stable reference such as a frequency comb, or by replacing it with an intrinsically more stable MIR source such as a CO$_2$ laser. 

\section{\label{Methods} Methods}

\subsection{\label{Meth-1} Devices structure and fabrication}

A 100nm-thick, lattice-matched Ga$_{0.51}$In$_{0.49}$P etch-stop layer followed by the Al$_{0.2}$Ga$_{0.8}$As/GaAs heterostructure is grown by MBE on top of a semi-insulating GaAs substrate. The heterostructure is sandwiched between 50 and 100nm-thick top and bottom n-doped contact layers with concentrations $3 \times 10^{18}$cm$^{-3}$ and $4 \times 10^{11}$cm$^{-3}$, and consists of six, 6nm-thick GaAs QWs with the central 5nm $n$-doped at $6 \times 10^{17}$cm$^{-3}$, separated by 40nm-thick, undoped Al$_{0.2}$Ga$_{0.8}$As barriers. 

The epi-layer is first transferred onto a 2”-diameter high-resistivity Si wafer using Au–Au thermo-compression bonding. The fabrication begins by wet etching the GaAs substrate and the etch-stop layer. Next, a Ti/Au (8nm/300nm) top Schottky contact is realized through e-beam lithography, followed by e-beam evaporation and lift-off. The epi-layers are subsequently ICP etched using the top metal layer as etch-mask. The ground metal layer is dry-etched by an Ar+ ion-beam around the patch-antenna matrix down to the Si substrate. A 100-nm-thick Si$_{3}$N$_{4}$ layer is then deposited on the Si by plasma enhanced chemical vapor deposition.
To electrically connect the patch-antennas, suspended $\sim$150-nm-wide Ti/Au (20nm/600nm) wire-bridges are fabricated by a two-step e-beam lithography process. A first resist layer is used as support after deposition, e-beam lithography and reflow, followed by a second one to define the wires by standard lift-off process. The same process is used to realize the air-bridge connecting the 2D array to the 50$\Omega$ coplanar line. The latter is deposited on the Si$_{3}$N$_{4}$ to prevent current leakage between the line's electrodes and the Si substrate.

\subsection{\label{Meth-3} Derivation of the electrical frequency response}

If $P_{1}$ and $P_{2}$ are the incident powers generated by the two QCLs, the total optical power incident on the biased photo-conductor is given by:
\begin{equation}
\label{E1}
P(t) = P_{tot}[1+m \cdot sin(\omega)t],
\end{equation}
where $P_{tot} = P_{1}+P_{2}$, $\omega$ is the difference between the two optical frequencies , and $m=2\sqrt{P_{1}P_{2}}/P_{tot}$ is the modulation index. If $R$ is the photodetector responsivity, the generated photocurrent $I_{ph}(t) = R \cdot P(t)$ can be split into a $dc$ component $I_{dc}=R \cdot P_{tot}$, which corresponds to the measured $dc$ photocurrent, and an $ac$ component of amplitude $I_{ac}=m \cdot R \cdot P_{tot}= m \cdot I_{dc}$.  In the absence of a sizeable resistance in series with the QWIP active region, as is the case here, it can be shown  that the amplitude of the current source $I_{s}$ in the photodetector small signal equivalent circuit (\textcolor{blue}{Supplement 1}) is precisely equal to $I_{ac} \simeq I_{dc}$  (since $m\simeq 1$ for the powers used in this work) \cite{Hakl2021,Peytavit2021}. 
The electrical frequency response of the QWIP is then obtained from the expression of the average $ac$ power dissipated in the $R_{L}=50\Omega$ input impedance of the microwave power-meter: 
\begin{equation}
\label{E2}
P_{L}(\omega) = \frac{1}{2}I_{dc}^{2}\bigg|\frac{R}{R+(R_{L}+i\omega L)(1+i\omega RC)}\bigg|^{2}R_{L}.
\end{equation}

To match quantitatively the power levels obtained experimentally in Fig.~\ref{fig:Response} in the main text, we used an amplitude of the $ac$ current source $I_{s} = I_{dc}/2 $ where $I_{dc}$ is the experimental $dc$ photocurrent generated by the two QCLs ($I_{dc} = 4.1$mA, $2.8$mA, and $1.25$mA for devices M5, M3 and M2 respectively). However, as discussed above, ideally we would rather expect $I_{s} = I_{dc}$, i.e. the generated heterodyne power should be $\sim 4$ times higher than what found experimentally. At the moment, we don’t have a clear explanation for this discrepancy, that could be in part attributed to a partial saturation of the ISB transition, each time the incident optical power oscillating at the difference frequency between the two QCLs reaches its maximum. Further measurements will be needed to validate this hypothesis. 

\subsection{\label{Meth-4} Comparison of single-shot acquisition and pressure detection limit}

\textcolor{black}{In Fig.M1 we report the absorption dip in the time domain corresponding to the saQ(3,3) transition at a nominal pressure of 10Pa, obtained from the chirped pulse (panel (a)) and from the QWIP current pulse (panel (b)). The black lines were recorded in single-shot, while the red one was obtained with 100 averages (same averaging used for Fig.~\Ref{fig:NH3-Spectroscopy-1}(b)). The SNRs in single-shot from the chirped and current pulse and  are respectively $\sim 8$ and $2$. From these numbers, based on the transmission intensities computed with Spectracalc\textsuperscript{\textregistered}, we estimate, for our 8cm-long gas cell, a minimum detectable gas pressures in single-shot of $\sim 0.3$Pa and $\sim 1.2$Pa.} 

\begin{figure}
\includegraphics[width=\linewidth]{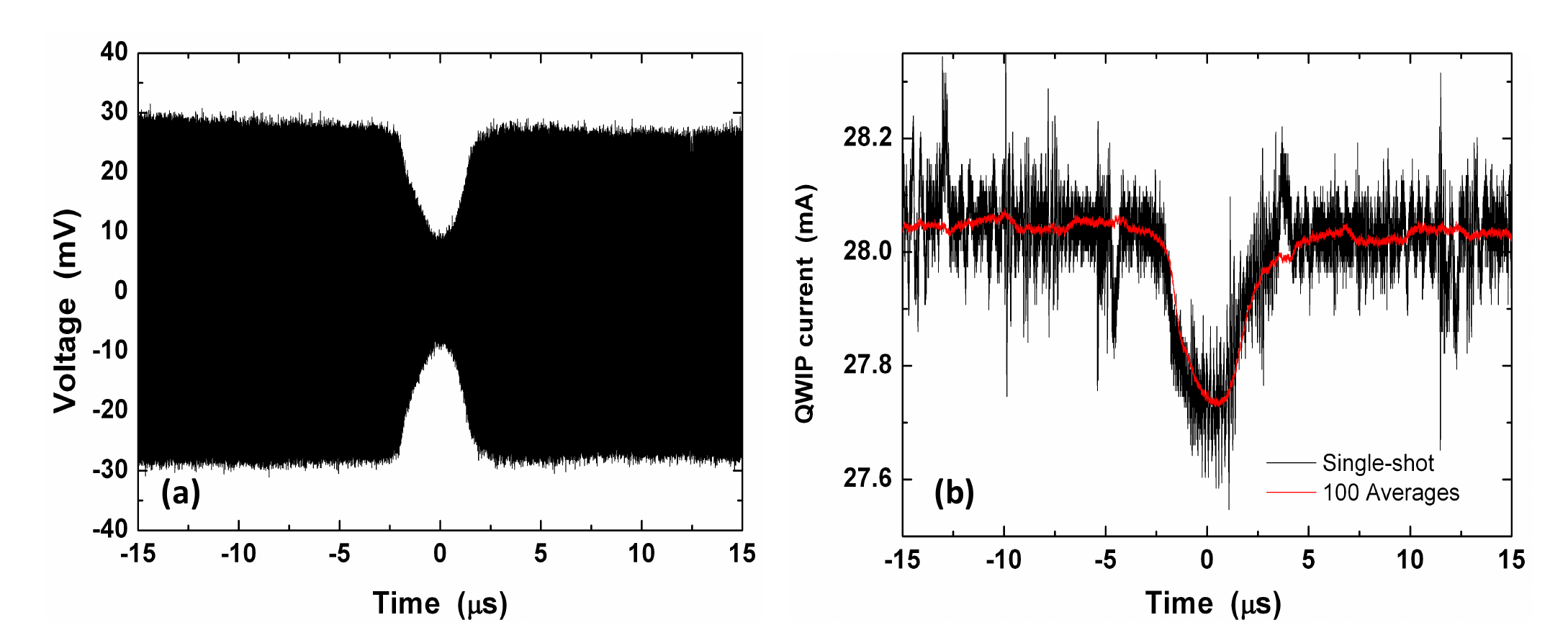}
			\textcolor{black}{Fig. M1. Absorption dip in the time domain corresponding to the saQ(3,3) transition at a nominal pressure of 10Pa. (a) Chirped pulse in single-shot. (b) QWIP current pulse in single-shot (black) and with 100 averages (red).}
				\label{Single-shot}
\end{figure}

\subsection{\label{Meth-2} Voltage source response time}

In Fig.~\ref{fig:NH3-Spectroscopy-2}, the three highest frequency NH$_{3}$ transition lines of the red spectrum (derived from the current pulse) present a systematically higher transmission compared to the green one (derived from the heterodyne pulse), \textcolor{black}{as well as a slight line asymmetry}. 
We attribute these facts to the finite transient response time, of approximately 30$\mu$s, of the voltage source  used to bias the QCL (Keithely 2440 5A SourceMeter). Indeed, from longer to shorter times (i.e. from lower to higher absolute frequencies in Fig.~\ref{fig:NH3-Spectroscopy-2}) the increase of the frequency chirp (see Fig.~\ref{fig:ChirpedPulse}(b)), leads to progressively temporally narrower transmission dips as shown in Fig.~\ref{fig:NH3-Spectroscopy-1}(b). 
As a result, at some point the rise time associated to a given transition becomes too short compared to the time needed by the voltage source to change its current in order to maintain a constant bias across the QWIP. Eventually this fact prevents reaching the transmission minimum. This is clearly the case for the highest frequency transition (i.e. the temporally narrowest), for which the associated rise time is of only $\sim 10 \mu$s, contrarily to the $\sim 100\mu$s of the lowest frequency one. Such experimental artifact is not present on the $ac$ port of the bias-tee, where slow bias variations are filtered out.
\\
\begin{backmatter}

\bmsection{Acknowledgments}
We gratefully acknowledge Raffaele Colombelli for helpful discussions on intersubband saturation and Etienne Okada for technical support during the RF measurements.

\bmsection{Funding} ANR Project Hispanid; RENATECH (French Network of Major Technology Centres); Project COMPTERA - ANR 22-PEEL-0003; Contrat de Plan Etat-Region (CPER) WaveTech. Wavetech is supported by the Ministry of Higher Education and Research, the Hauts-de-France Regional council, the Lille European Metropolis (MEL), the Institute of Physics of the French National Centre for Scientific Research (CNRS) and the European Regional Development Fund (ERDF).

\bmsection{Disclosures} The authors declare no conflicts of interest.

\bmsection{Supplemental document}
See Supplement 1 for supporting content. 

\end{backmatter}
\bibliography{Lin_et_al}

%Manual citation list
%\begin{thebibliography}{1}
%\bibitem{Zhang:14}
%Y.~Zhang, S.~Qiao, L.~Sun, Q.~W. Shi, W.~Huang, %L.~Li, and Z.~Yang,
 % \enquote{Photoinduced active terahertz metamaterials with nanostructured
  %vanadium dioxide film deposited by sol-gel method,} Opt. Express \textbf{22},
  %11070--11078 (2014).
%\end{thebibliography}

\end{document}